\documentclass{natureprintstyle}

\spacing{1}
\spacing{2}
\spacing{1}

\usepackage{amsmath,amssymb}
\usepackage{graphicx}
\usepackage{color}
\usepackage{bm}
\usepackage{longtable}
\usepackage{amssymb}
\usepackage{rotating}
\usepackage{hyperref}

\usepackage{hyperref}

\let\citep\cite
\let\citet\cite

\title{\sffamily Strongly Lensed Repeating Fast Radio Bursts as Precision Probes of the Universe}

\author{\sffamily Zhengxiang Li$^{1}$, He Gao$^{1}$, Xuheng Ding$^2$, Guojian Wang$^{1}$, Bing Zhang$^{3,4,5}$}

\begin{document}
\maketitle

\begin{affiliations}
\item Department of Astronomy, Beijing Normal University, Beijing 100875, China	
\item School of Physics and Technology, Wuhan University, Wuhan 430072, China
\item Department of Physics and Astronomy, University of Nevada, Las Vegas, NV 89154, USA
\item National Astronomical Observatories of China, Chinese Academy of Sciences, Beijing 100012, China
\item Department of Astronomy, School of Physics and Kavli Institute for Astronomy and Astrophysics, Peking University, Beijing 100871, China
\end{affiliations}

\let\thefootnote\relax\footnote{Corresponding author: H. Gao: gaohe@bnu.edu.cn}

\vspace{-0.5mm}
\begin{abstract}
\sffamily
Fast Radio bursts (FRBs) are bright transients with millisecond durations at $\sim$ GHz frequencies and typical redshifts $z$ probably $>0.8$. They are likely to be gravitationally lensed by intervening galaxies. Since in a strongly lensed FRB system, the time delay between images can be measured to extremely high precision because of the large ratio $\sim10^9$ between the typical galaxy-lensing delay time $\sim\mathcal{O}$(10 days) and the narrow width of the bursts $\sim\mathcal{O}$(ms), we propose accurate measurements of time delays between images of lensed FRBs as precision probes of the universe. Here we show that, within the flat $\Lambda$CDM model, the Hubble constant $H_0$ can be constrained with a sub-percent level ($\sim0.91\%$) uncertainty from 10 such systems, which could be observed with the Square Kilometer Array (SKA) within $<$ 30 years. More importantly, the cosmic curvature can be constrained to a precision of $\sim0.076$ in a model-independent manner. Such a direct and model-independent constraint on the cosmic curvature will provide a stringent direct test for the validity of the cosmological principle and break the intractable degeneracy between the cosmic curvature and dark energy.
\end{abstract}

\sffamily

\begin{center}
{\textbf{ \Large \uppercase{Introduction}} }
\end{center}

	Fast Radio bursts (FRBs) are bright transients with millisecond durations at $\sim$ GHz frequencies, whose physical origin is subject to intense debate \cite{lorimer07,thornton13}. Most FRBs are located at high galactic latitudes and have anomalously large dispersion measures (DMs). Attributing DM to an intergalactic medium origin, the corresponding redshifts $z$ are typically $>0.8$. Up to now, more than 30 FRBs have been published \cite{petroff16}. One of them, FRB 121102, shows a repeating feature \cite{spitler16}. The repetition of FRB 121102 enables high-time-resolution radio interferometric observations to directly image the bursts, leading to the localization of the source in a star-forming galaxy at $z=0.19273$ with sub-arcsecond accuracy \cite{chatterjee17}. The cosmological origin of the repeating FRBs is thus confirmed. For other FRBs, although no well-established evidence being published, they are also strongly suggested to be of a cosmological origin, due to their all-sky distribution and their anomalously large values of dispersion measures (DM) \cite{thornton13}. It is possible that all FRBs might be repeating, and only the brightest ones are observable. On the other hand, it is also possible that repeating and non-repeating FRBs may originate from different progenitors \cite{palaniswamy17}. Interestingly, these transient radio sources are likely lensed from small to large scales, e.g., through plasma lensing in their host galaxies \cite{cordes17}, gravitational microlensing by an isolated and extragalactic stellar-mass compact object \cite{zheng14, munoz16}, and strong gravitational lensing by an intervening galaxy \cite{li14, dai17}. Here we only focus on the possibility of strongly gravitationally lensed FRBs and their applications to conduct cosmography. Therefore, in our following analysis, ``lensed FRBs" only refers to the case that an FRB is strongly gravitationally lensed by an intervening galaxy. For a lens galaxy with the mass of dark matter halo  $\sim 10^{12}M_{\odot} h^{-1}$ ($h$ is the Hubble constant in units of 100 ~$\rm{km~s^{-1}~Mpc^{-1}}$), the typical time delay and angular separation between different images of lensed FRBs are $\sim\mathcal{O}$(10 days) and of the order arcseconds, respectively. These multiple images of lensed FRBs cannot be resolved by radio survey telescopes since  their typical angular resolution is of the order of 10 arcminutes. Therefore, a lensed non-repeating FRB source may be observed as a repeating FRB source, showing two to four bursts with respective time delays of several days. The DM value and the scatter-broadening of each burst could be slightly or even significantly different from each other depending on the plasma properties along different lines of sight (LOS). It is therefore difficult to identify the lensed non-repeating FRBs. However, if a repeating FRB is strongly lensed by an intervening galaxy, a series of image multiplets from the same source will exhibit a fixed pattern in their mutual time delays, appearing over and over again as we detect the repeating bursts \cite{dai17}. Observations of FRB 121102 in radio and its counterpart in optical indicate that this repeater is not lensed  (no intervening lens galaxy or multiple images of the host are observed) and the intrinsic repetition happens randomly. Therefore, a fixed temporal pattern associated with a future repeating FRB source would be a smoking-gun signature that it is strongly-lensed. Each burst emitted from the source would travel through different paths to reach to the observer with time delays. If these lensed bursts can be imaged, they should appear as different images in the sky. Their spectra and lightcurves might be slightly different from each other because of different paths they traveled through, so that the morphology of bursts may not be the main feature to identify lensed FRBs. For a series of randomly-generated repeating bursts, the intrinsic time difference between each two adjacent bursts should be the same for all lensed (two or four) images. Therefore, a fixed time pattern of all the repeating bursts is the most robust evidence for identifying a lensed FRB system. Once a survey telescope registered a fixed time pattern repeating two or four times with a delay $\sim\mathcal{O}$(10 days), one could then employ more powerful radio telescopes such as Very Large Array (VLA) or the future SKA to observe more repetitions and resolve multiple images of the bursts. Meanwhile, one could observe the source using optical and near-IR telescopes to identify an intervening lens galaxy near the LOS as well as the multiple images of the host galaxy (Einstein ring or arcs) with angular separations of the order of arcseconds. If the image locations in both radio and optical (or near-IR) bands match each other, in combination with the fixed time delay repetition pattern mentioned above, a lensed FRB system can be confirmed and the host and the lens galaxies identified.
	
	Current FRB observations suggest a sufficiently high all-sky FRB rate of $\sim10^{3}-10^{4}$ per day \cite{thornton13,champion16}. Upcoming surveys  such as the Swinburne University of Technology's digital backend for the Molonglo Observatory Synthesis Telescope array (UTMOST) \cite{caleb16}, the Hydrogen Intensity and Real-time Analysis eXperiment (HIRAX) \cite{newburgh16}, the Canadian Hydrogen Intensity Mapping Experiment (CHIME) \cite{bandura14}, and especially the SKA project \cite{macquart15} will map a considerable fraction of the sky with a detection rate of FRBs of $>100$ per day \cite{fialkov17}. For an FRB happening at $z\gtrsim1$, the probability for it to be strongly lensed is $\sim$ a few$\times10^{-4}$ \cite{hilbert08}. As a result, future radio surveys, such as SKA, will have the ability to discover $>10$ strongly lensed FRBs per year \cite{li14, munoz16, dai17}. According to the current data, at least $3\%$ ($1/30$) observed FRBs are repeating FRBs. With a conservative estimate, $\sim$10 strongly lensed repeating FRBs are expected to be accumulated within $<30$ years with the operation of SKA. 
	
	Owing to the small ratio ($\sim10^{-9}$) between the the short duration of each burst $\sim\mathcal{O}$(ms) and the typical galaxy-lensing delay time $\sim\mathcal{O}$(10 days), time delays between images of these systems can be measured to great precision. Moreover, due to overwhelmingly accurate localizations of lensed FRB images from deep VLA observations (or future SKA observations) and clean high-resolution images of the host galaxy without a dazzling active galactic nucleus (AGN), the mass profile of the lens can be also modeled with high precision. Therefore, we propose that lensed FRB systems can be a powerful probe for studying cosmology. Lensing theory predicts that the difference in arrival time between image A and image B, i.e. the ``time delay" $\Delta\tau_{\rm AB}$, is expressed as
	\begin{equation}\label{eq1}
	\Delta\tau_{\rm AB}=\frac{\Delta \Phi_{\rm AB}}{c}D_{\Delta t}=\frac{\Delta \Phi_{\rm AB}}{c}\cdot(1+z_{l})\frac{D^{\rm A}_{l}D^{\rm A}_{s}}{D^{\rm A}_{ls}},
	\end{equation}
	where $\Delta \Phi_{\rm AB}$ is the Fermat potential difference between the two image positions, $c$ is the speed of light, $D_{\Delta t}$, the so-called ``time delay distance", is just a multiplicative combination of the three angular diameter distances ($D^{\rm A}_{l}$: from the observer to the lens, $D^{\rm A}_{s}$: from the observer to the source, $D^{\rm A}_{ls}$: from the lens to the source), $z_l$ is the redshift of lens. This quantity has the dimension of distance and is inversely proportional to the Hubble constant, $H_0$, which sets the age and length scale for the present universe and is one of the most important parameter for cosmology. Therefore, the time delay distance $D_{\Delta t}$ is primarily sensitive to $H_0$ and that measured from lensed quasar systems has been used to measure the Hubble constant. Moreover, the relations among these three angular diameter distances is highly dependent on the geometric properties of the space. We introduce dimensionless comoving angular diameter distances, $d_{l}\equiv d(0, z_l)\equiv (1+z_l)H_0D^{\rm A}_{l}/c$, $d_{s}\equiv d(0, z_s)\equiv (1+z_s)H_0D^{\rm A}_{s}/c$, and $d_{ls}\equiv d(z_l, z_s)\equiv (1+z_s)H_0D^{\rm A}_{ls}/c$ ($z_s$ is the redshift of source), to illustrate this. For example, qualitatively and intuitively, $d_{\rm s}$ is greater, equal to, or smaller than $d_{\rm l}+d_{\rm ls}$ if the space is open, flat, or closed, respectively (see Methods section for quantitative details). Therefore, in combination with distances from type Ia supernova (SNe Ia) observations, the time delay distance can be used to directly measure the spatial curvature $\Omega_k$ in a cosmological-model-independent manner. 
	Decades of observations have ushered in the era of precision cosmology. The flat $\Lambda$CDM model is found to be consistent with essentially all the conservational constraints. Yet recent direct local-distance-ladder measurements of $H_0$ have reached a $2.4\%$ precise measurement: $H_0=73.24\pm1.74~\rm{km~s^{-1}~Mpc^{-1}}$ \cite{riess16}, which greatly increased the tension with respect to the latest $Planck$-inferred value ($H_0=67.27\pm0.66~\rm{km~s^{-1}~Mpc^{-1}}$) \cite{planck15} to ~3.4$\sigma$. Lensed FRBs, as a powerful probe and completely independent dataset based on a different physical phenomenon, would provide complementary information and therefore are of vital importance to clarify this issue.

\begin{center}
{\textbf{ \Large \uppercase{Results}} }
\end{center}

\begin{figure}[t]
	\centering
	\includegraphics[width=0.47\textwidth, height=0.29\textwidth]{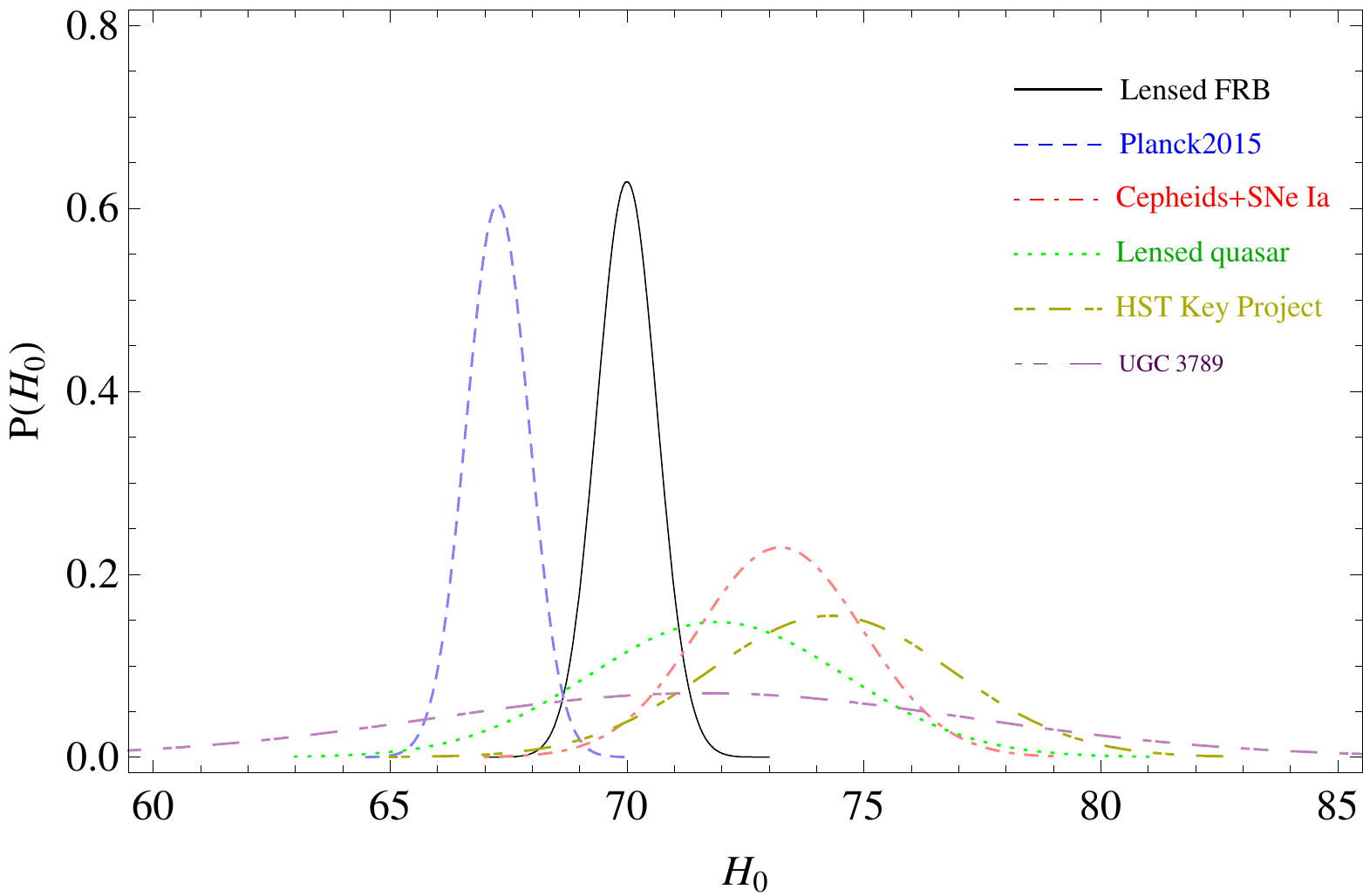}\\
	\caption{Probability distribution functions (PDFs) of the Hubble constant constrained from 10 lensed FRBs and some other currently available observations. Besides the result obtained from 10 lensed FRB systems in this work (the black solid line), from top to bottom the lines represent $H_0$ inferred from the $Planck$ satellite CMB measurements  ($67.27\pm0.66~\rm{km~s^{-1}~Mpc^{-1}}$) \citep{planck15},  local distance measurements ($73.24\pm1.74~\rm{km~s^{-1}~Mpc^{-1}}$) \citep{riess16}, time-delay cosmography of strongly lensed quasars ($71.9^{+2.4}_{-3.0}~\rm{km~s^{-1}~Mpc^{-1}}$) \citep{suyu17}, distance measurements from the Hubble Space Telescope (HST) key project ($74.3\pm2.6~\rm{km~s^{-1}~Mpc^{-1}}$) \citep{freedman12}, and VLBI observations of water masers orbiting within the accretion disc of UCG 3789 ($71.6\pm5.7~\rm{km~s^{-1}~Mpc^{-1}}$) \citep{reid12}, respectively. }\label{fig1}
\end{figure}

	In order to investigate the constraining power of lensed FRBs on some fundamental cosmological parameters, we perform a series of simulations with the proper inputs in the following three aspects (see Methods section for details):  i) the redshift distribution of incoming FRBs; ii) for a source at redshift $z_s$, the lens redshift $z_l$ to produce the maximal differential lensing probability;  iii) the uncertainty of each factor contributing to the accuracy of time delay distance measurement. Since the time delay distance is very sensitive to the Hubble constant, we estimate the constraining power on $H_0$ by simulating 10 lensed FRBs in the flat $\Lambda$CDM with the matter density being fixed as $\Omega_m=0.3$. With the assumed fiducial model (flat $\Lambda$CDM model with $H_0=70~\rm{km~s^{-1}~Mpc^{-1}}$ and $\Omega_m=0.3$) and three factors outlined above specified, we perform 10000 simulations each containing 10 lensed FRB systems and obtain the probability distribution of the estimated $H_0$. Two different redshift distributions, $N_{\rm const}(z)$ and $N_{\rm SFH}(z)$, are considered (see Methods section for details). They do not lead to significant differences in the constraint on $H_0$ and consistently give stringent constraints with a $\sim0.91\%$ uncertainty. Results are shown in Figure \ref{fig1}. It is suggested that compared to the currently available results, $\sim 10$ lensed FRBs will have obvious predominance in precision in constraining $H_0$. For instance, it improves by a factor $\sim5$ with respect to the current state-of-the-art case of lensed quasars \cite{suyu17}.

\begin{figure}[t]
		\centering
		\includegraphics[width=0.47\textwidth, height=0.29\textwidth]{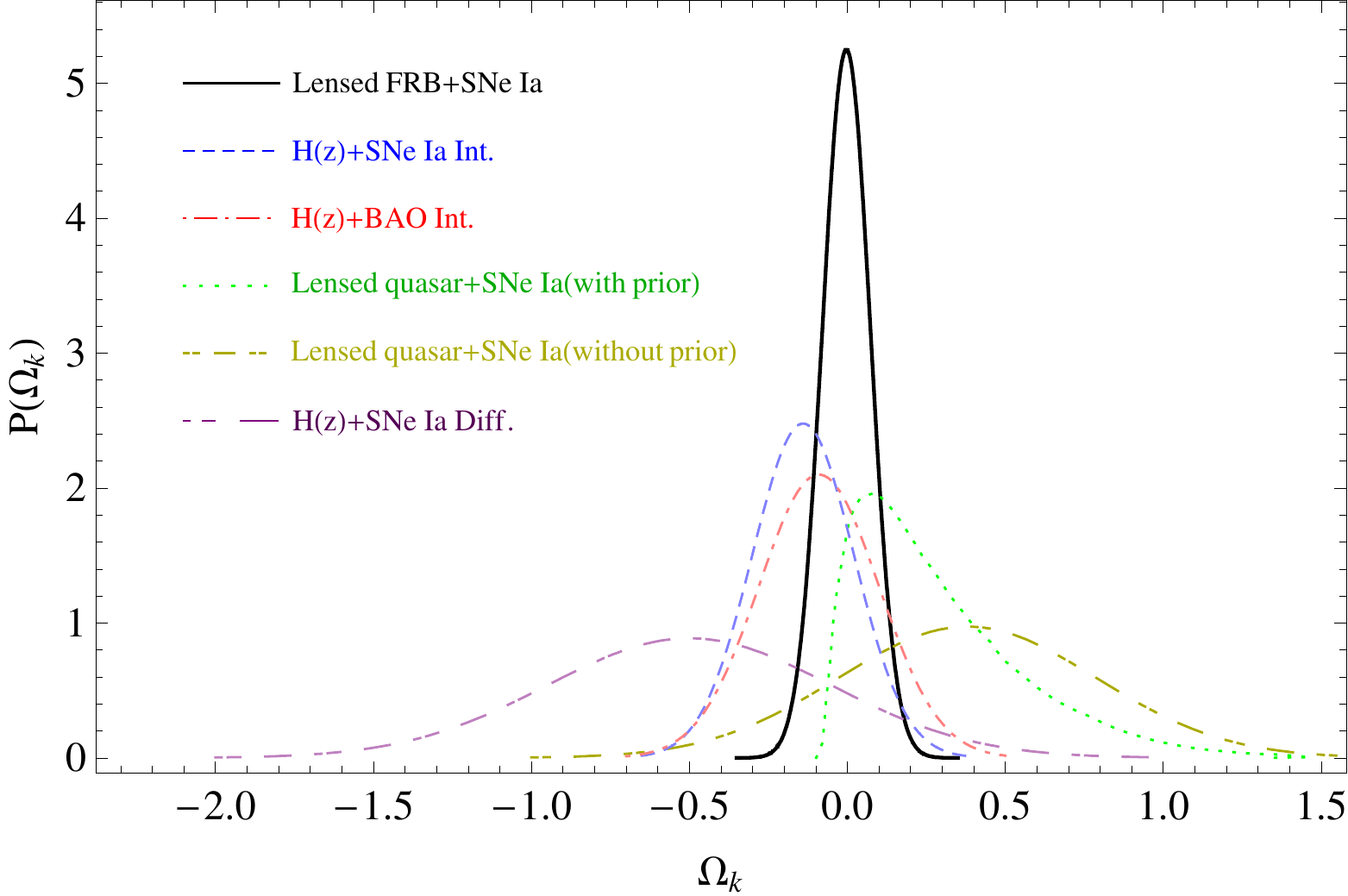}\\
		\caption{Model independent probability distribution functions (PDFs) of the cosmic curvature estimated from 10 lensed FRBs and some other currently available observations. Besides the result obtained from 10 lensed FRB systems in this work (the black solid line), from top to bottom the lines are $\Omega_k$ inferred from the integral method with expansion rate (i.e., the Hubble parameter $H(z)$) and SNe Ia observations ($-0.140\pm0.161$) \citep{li16}, the integral method with expansion rate and BAO observations ($-0.09\pm0.19$) \citep{yu16}, distance sum rule with the prior $\Omega_k>-0.1$ ($0.25^{+0.72}_{-0.33}$ ) \citep{rasanen15}, distance sum rule without the prior $\Omega_k>-0.1$ ($-0.38_{-0.84}^{+1.01}$) \citep{rasanen15}, and the differential approach with the expansion rate and SNe Ia observations ( $-0.50^{+0.54}_{-0.36}$) \citep{edvard11}, respectively. }\label{fig2}
	\end{figure}

	In addition to constraining the Hubble constant within the flat $\Lambda$CDM model, one can also give a model-independent estimate of the cosmic curvature using lensed FRBs. The spatial curvature of the universe is one of the most fundamental parameters. On one hand, estimating the curvature is a robust way to test the assumption that the universe is exactly described by the homogeneous and isotropic Friedmann-Lema\^{i}tre-Robertson-Walker (FLRW) metric \cite{clarkson08}. On the other hand, the spatial curvature is also closely related to some other important problems such as the evolution of the universe and the nature of dark energy \cite{clarkson07}.  Recently, the sum rule of distances along null geodesics of the FLRW metric was put forward as a consistency test for the validity of the homogeneous and isotropic background  \cite{rasanen15}. More recently, with an upgraded distance sum rule, time delay distance measurements from lensed quasars were proposed to test the FLRW metric and estimate the cosmic curvature (see Methods section for details) \cite{liao17a}. Here in combination with $\sim4000$ type Ia supernova observations from the near future Dark Energy Survey (DES), we examine the ability of the lensed FRBs for constraining the cosmic curvature. We find that the constraints from lensed FRBs with the two considered redshift distributions are very similar and the spatial curvature parameter can be constrained to a precision of $\sim0.076$. Results from lensed FRBs and other currently available model-independent methods are presented in Figure \ref{fig2}. Again, in this model-independent domain, lensed FRBs are the most promising tools for constraining the cosmic curvature. Moreover, the precision of the results from lensed FRBs potentially approaches that inferred from the $Planck$ satellite observations within the standard $\Lambda$CDM model, where $\Omega_k=-0.004\pm0.015$ was obtained \cite{planck15}.

\begin{center}
{\textbf{ \Large \uppercase{Discussion}} }
\end{center}

	Here we propose strongly lensed repeating FRBs as a precision cosmological probe. Representatively, we investigate the constraining power of lensed FRB systems observed in the near future on two of the most important cosmological parameters, Hubble constant $H_0$ and cosmic curvature $\Omega_k$. For $H_0$, we obtain that it can be constrained with a relative $\sim0.91\%$ uncertainty from 10 lensed FRB systems. This promising constraint with sub-percent uncertainty level suggests that lensed FRBs, as a powerful probe and completely independent dataset based on a different physical phenomenon, would provide complementary information and therefore are of vital importance to clarify the tension between the latest $Planck$-inferred $H_0$ and the one from direct local-distance-ladder observations. For $\Omega_k$, it can be constrained to a precision of $\sim0.076$ from 10 lensed FRB systems in a model-independent way on the basis of the distance sum rule. This result is the most precise one in the field of model-independent estimations for the cosmic curvature. On one hand, such a direct and model-independent constraint on the cosmic curvature will provide a stringent direct test for the validity of the Friedmann-Lema\^{i}tre-Robertson-Walker metric and break the intractable degeneracy between the cosmic curvature and dark energy, offering the opportunity in investigating the nature of dark sectors of the universe. On the other hand, having model-dependent and direct measurements of the same quantity is of utmost importance. In the absence of significant systematic errors, if the standard cosmological model is the correct one, indirect (model-dependent) and direct (model-independent) constraints on this parameter should be consistent. If they were significantly inconsistent, this would provide evidence of physics beyond the standard model or unaccounted systematic errors. Strongly lensed FRBs can help to reach such a goal.

\begin{center}
{\textbf{ \Large \uppercase{Methods}} }
\end{center}

	In order to examine the potential of using lensed FRBs as cosmological probes, three related aspects need to be addressed: i) the redshift distribution of the incoming FRBs; ii) for a source at redshift $z_s$, the lens redshift $z_l$ to produce the maximal differential lensing probability;  iii) the uncertainties of different factors contributing to the accuracy of time delay distance measurements. We discuss these three items one by one. In addition, we also introduce the distance sum rule for estimating the cosmic curvature.

\noindent
{\textbf{FRB redshift distribution}}
    We consider two possible scenarios suggested in \cite{munoz16}. The first one invokes a constant comoving number density, so that the number of FRBs in a shell of width $dz$ at redshift $z$ is proportional to the comoving volume of the shell $dV(z)$ \cite{oppermann16}. By introducing a Gaussian cutoff at some redshift $z_{\rm{cut}}$ to represent an instrumental signal-to-noise threshold, the constant-density distribution fuction $N_{\rm{const}}(z)$ is expressed as
    \begin{equation}
    N_{\rm{const}}(z)=\mathcal{N}_{\rm{const}}\frac{\chi^2(z)}{H(z)(1+z)}e^{-{D^{\rm{L}}}^2(z)/[2{D^{\rm{L}}}^2(z_{\rm cut})]},
    \end{equation}
    where $\chi(z)$ is the comoving distance and $D^{\rm{L}}$ is the luminosity distance. $\mathcal{N}_{\rm{const}}$ is a normalization factor to ensure that the integration of  $N_{\rm{const}}(z)$ is unity and $H(z)$ is the Hubble parameter at redshift $z$. The second distribution requires that FRBs follow the star-formation history (SFH) \cite{caleb16}, so that
    \begin{equation}
    N_{\rm{SFH}}(z)=\mathcal{N}_{\rm{SFH}}\frac{\dot{\rho}_*(z)\chi^2(z)}{H(z)(1+z)}e^{-{D^{\rm{L}}}^2(z)/[2{D^{\rm{L}}}^2(z_{\rm cut})]},
    \end{equation}
    where $\mathcal{N}_{\rm{SFH}}$ is the normalization factor and is chosen to have $N_{\rm{SFH}}(z)$ integrated to unity. The density of star-formation history is parametrized as 
    \begin{equation}
    \dot{\rho}_*(z)=h\frac{\alpha+\beta z}{1+(z/\gamma)^\delta},
    \end{equation}
   with $\alpha=0.017,~\beta=0.13,~\gamma=3.3,~\delta=5.3$, and $h=0.7$ \cite{cole01, hopkins06}. For redshifts of currently available FRBs, different from previous estimation using a simple relataion between DM and $z$ proposed by Ioka \cite{Ioka03}, we re-estimate them with a more precise DM-$z$ relation given in \cite{deng14}. It is found that the inferred $z$ values are systematically greater than previously estimated ones, which are typically $>0.8$ and with several FRBs having $z >1$ even after properly subtracting the DM contribution from the FRB host. In this case, for these two FRB distribution functions, a cutoff $z_{\rm cut}=1$ is chosen to match redshifts of currently detected events. In our analysis, $N_{\rm const}$ and $N_{\rm SFH}$ are employed to investigate whether cosmological implications from lensed FRBs are dependent on the assumed redshift distributions.

\noindent
{\textbf{Lensing probability}}
According to the lensing theory \cite{schneider92}, the probability for a distant source at redshift $z_s$ lensed by an intervening dark matter halo is 
\begin{equation}
P=\int_0^{z_s}dz_l\frac{dD^{\rm p}}{dz_l}\int_0^\infty \sigma(M, z_{l})n(M, z_l)dM,
\end{equation}
where $dD^{\rm p}/dz_l$ is the proper distance interval, $\sigma(M, z_l)$ is the lensing cross-section of a dark matter halo with its mass and redshift being $M$ and $z_l$, respectively, $n(M, z_l)dM$ is the proper number density of the deflectors with masses between $M$ and $M+dM$. For a singular isothermal sphere (SIS) lens, the cross-section producing two images with a flux ratio being smaller than a given threshold $r$ is 
\cite{li02}
\begin{equation}
\sigma(<r)=16\pi^3\bigg(\frac{\sigma_v}{c}\bigg)^4\bigg(\frac{r-1}{r+1}\bigg)^2\bigg(\frac{D^{\rm A}_{l}D^{\rm A}_{ls}}{D^{\rm A}_{s}}\bigg)^2,
\end{equation} 
where $\sigma_v$ is the velocity dispersion. Moreover, the comoving number density of dark matter halos within the mass range ($M,~M+dM$) at redshift $z$ is 
\begin{equation}
n(M, z)dM=\frac{\rho_0}{M}f(M, z)dM,
\end{equation}
where $\rho_0$ is the present value of the mean mass density in the universe, and $f(M, z)dM$ is the Press-Schechter function \cite{press74}. For any FRB at redshift $z_s$ following the distribution $N_{\rm const}$ or $N_{\rm SFH}$, we determine the lens redshift $z_l$ by maximizing the differential lensing probability, $dP/dz$. Assuming an SIS-like lens halo of mass $M=10^{12}M_{\odot} h^{-1}$ ($h=H_0/100~\rm{km~s^{-1}~Mpc^{-1}}$) and $r\leq5$, the function of the lens redshift $z_l$ producing the maximal differential lensing probability with respect to the source reshift $z_s$ was shown in Figure 2 of \cite{li14}. In our analysis, this function is used to determine the lens redshift $z_l$ for any given source at redshift $z_s$.

\noindent
{\textbf{Uncertainty contribution}}
    In order to estimate the time delay distance from individual lensing systems for an accurate cosmography, as suggested in Equation~(\ref{eq1}),  it has been recognized that three key analysis steps should be carried out \cite{treu10, treu16}: i.e., 1) time delay measurement, 2) lens galaxy mass modeling which can be used to predict the Fermat potential differences, 3) and the line of sight (LOS) environment modeling, which is adopted to account for the weak lensing effects due to massive structures in the lens plane and along the LOS.

    The differences in the arrival time between images can be precisely measured for a lensed FRB system since the short duration of each burst $\sim\mathcal{O}$(ms) is much smaller ($\sim10^{-9}$) than the typical galaxy-lensing time delay $\sim\mathcal{O}$(10 days). Therefore, errors of time delay measurements for lensed FRBs are negligible. Compared to the best 3\% uncertainty of time delay measurements in traditional lensed quasars \cite{liao15, suyu17}, the precision for the case of FRBs is greatly improved. 

    For the lens galaxy mass modeling, it requires a high-resolution, good-quality image of the lensed host galaxy and accurate localizations of the lensed FRB images. The advantage of a lensed FRB system is that it does not have a bright AGN, so that clean host images can be obtained before or after FRB.  In practice, once a strongly lensed repeating FRB is identified by a large field-of-view radio survey program (e.g. CHIME), images of the lensed FRBs can be accurately localized from the deep follow-up observations with VLBI or SKA. High-quality optical images of the host galaxy can be obtained from follow-up facilities such as HST or the near future James Webb Space Telescope (JWST), which can be used to study the mass distribution of the deflector with lens modeling techniques.

\begin{figure*}[t]
     		\centering
     		\includegraphics[width=0.98\textwidth]{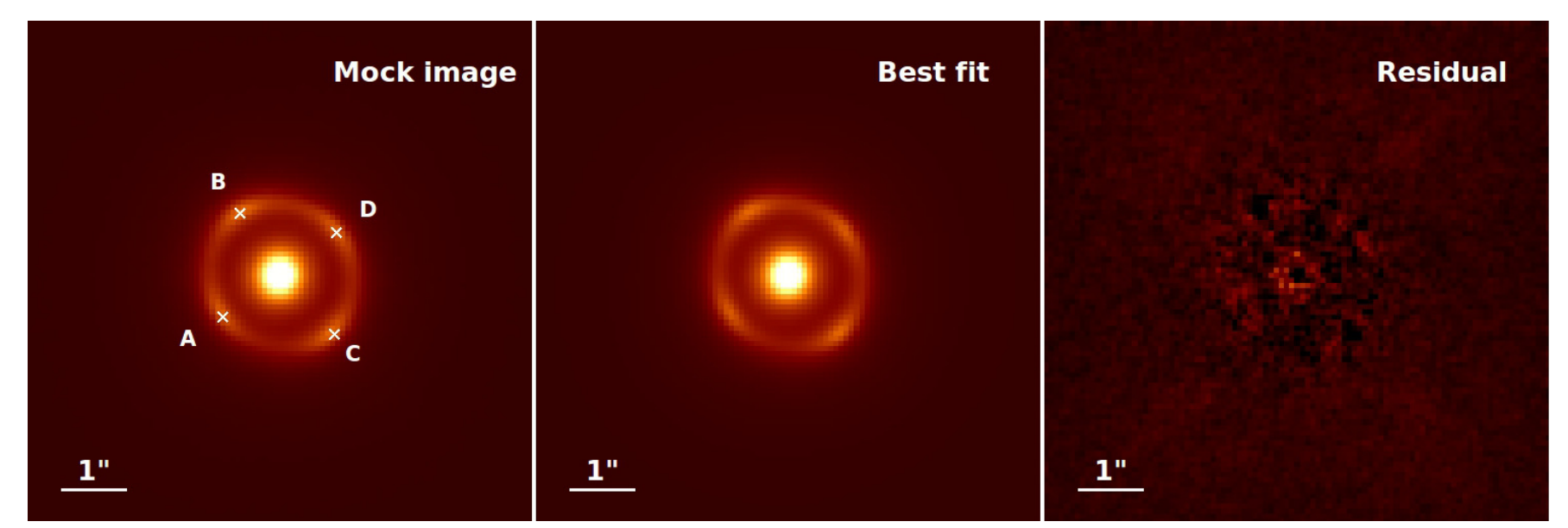}\\
     		\caption{{\bf Left:} Simulation results based on HST, WFC3/F160w with image drizzled to $0.08''$. {\bf Middle:} Best-fit image. {\bf Right:} Residual map.}\label{fig3}
     	\end{figure*}

\begin{figure}[t]
			\centering
			\includegraphics[width=0.5\textwidth]{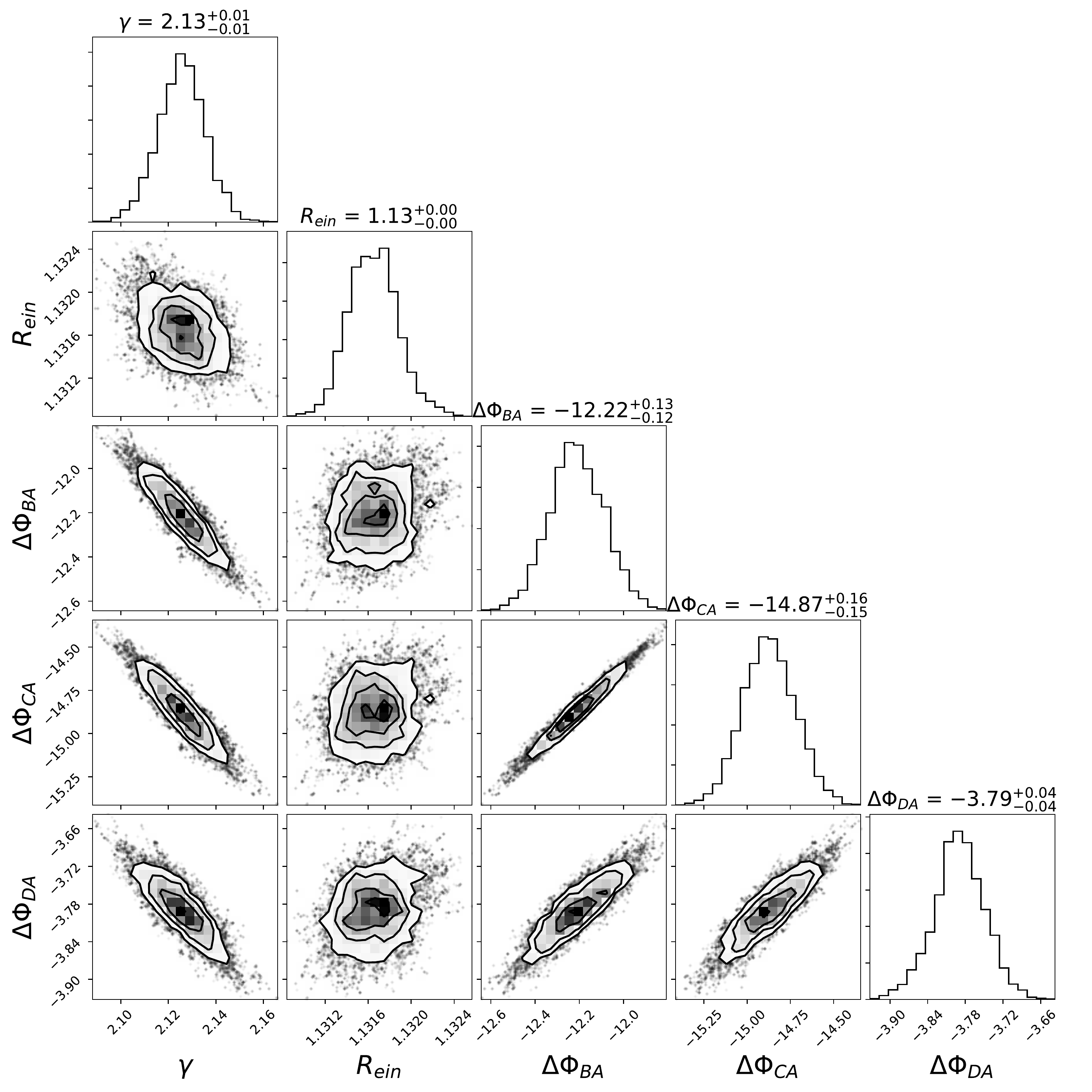}\\
			\caption{The contours of parameters inferred from the MCMC technique. The demonstrated parameters are the Einstein radius $R_{\mathrm{ein}}$ (in units of arcsecond), power-law mass profile slope $\gamma$ and differences of Fermat potentials between each pair of images. Note that Fermat potentials have not units, and we have re-scaled their values to better present the uncertainty level.}\label{fig4}
		\end{figure}

\begin{figure}[t]
			\includegraphics[width=0.45\textwidth]{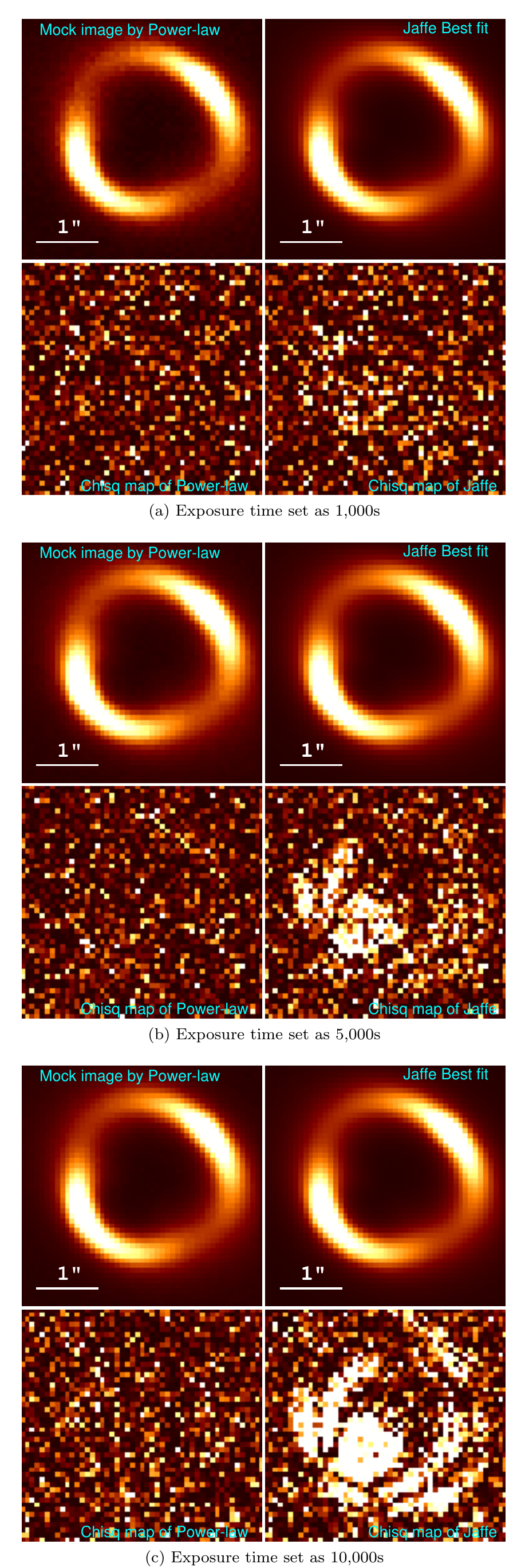}
			\caption{Results of the the simulation tests by generating the lensed arc with a power-law mass distribution model but fitting with the Jaffe model.}\label{fig5}
		\end{figure}

\begin{figure}[t]
			\centering
			\includegraphics[width=0.45\textwidth]{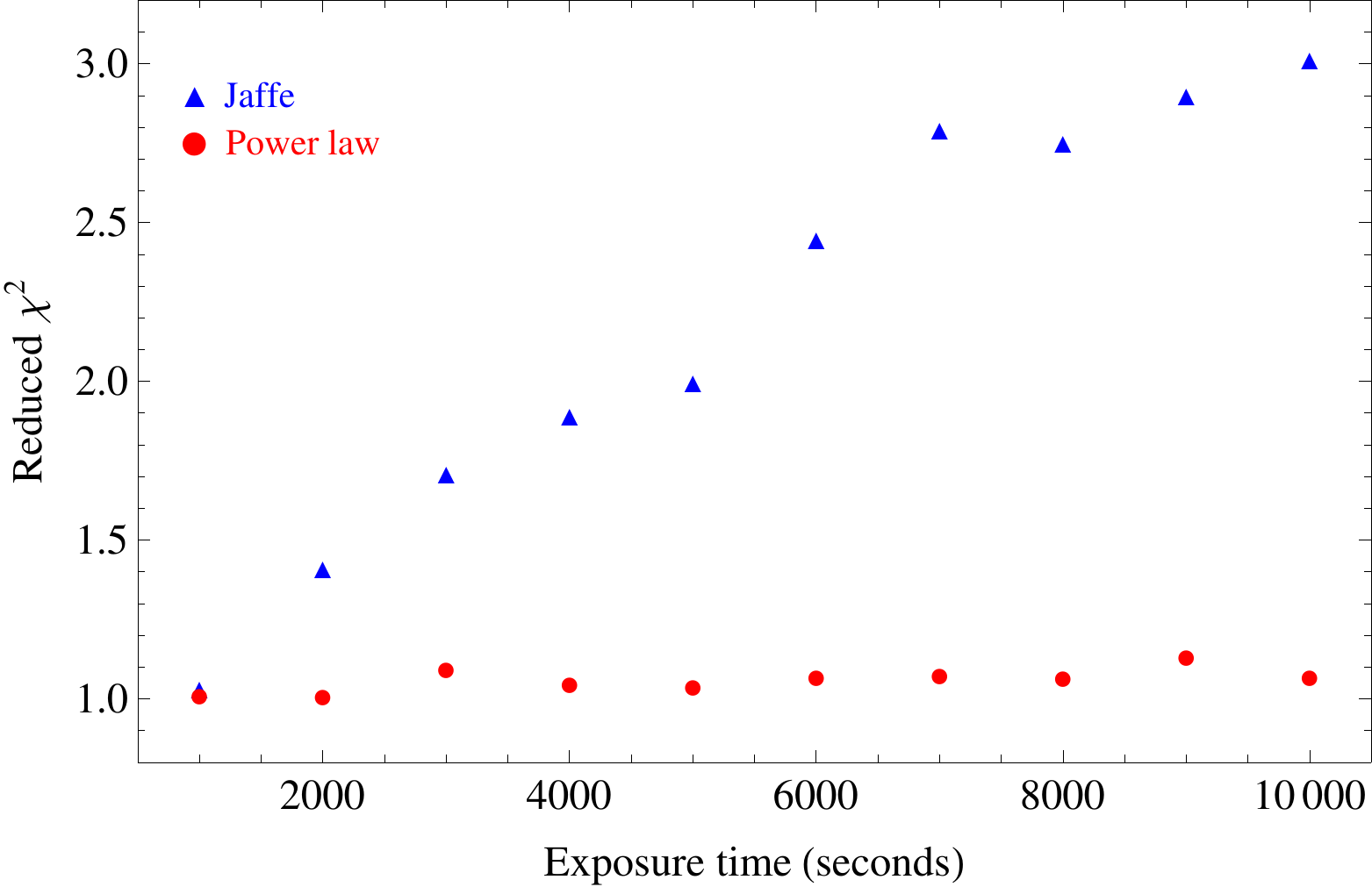}\\
			\caption{The reduced $\chi^2$ values by fitting the mock data with true model (i.e. power-law) and the wrong model (i.e. Jaffe) when varying the exposure time from 1,000s to 10,000s.}\label{fig6}
		\end{figure}

     In order to quantitatively estimate the uncertainty level of lens modeling from integrated lensed host image without a dazzling AGN, we carry out a series of simulations. First, we generate mock lensed images following the industrial standard as introduced in \cite{suyu17, ding17}. Specifically, in our simulation, the S\'ersic profile \cite{sersic63} is used to describe light profiles of the source (background) and the lens (foreground) galaxies. For lens mass profile, it is assumed to follow the power-law mass distribution of elliptical galaxies. Images are supposed to be observed by HST using the Wide Field Camera 3 (WFC3) IR channel in the F160W band. The settings related to the quality of mock images, such as the exposure time and drizzling process, are chosen based on the H0LiCOW program. Even though, in the simulation, FRB is non-luminous in optical/IR band and thus does not contribute any light to the surface brightness of images, locations of FRB images are considered to calculate the difference of the Fermat potential between each point source in the image plane. The final simulated image is shown in Fig.~\ref{fig3} (left panel). We apply a Monte Carlo Markov Chain (MCMC) approach to find the best fit parameters (and parameter uncertainty) for the light and mass profiles of the source and lens galaxies, by fitting the mock image with Glafic \cite{Oguri10}. The best-fit and the residual maps are shown in the middle and the right panel of Fig.~\ref{fig3}, respectively. We then calculate the differences of Fermat potentials between each pair of images based on the fitting results and plot their contours together with the slope of the power-law lens mass profile and the Einstein radius $R_{\mathrm{ein}}$ (see Fig.~\ref{fig4}). There appears to be an obvious degeneracy between the slope of the power-law mass profile $\gamma$ and the Fermat potential differences ($\Delta\Phi_{\mathrm{BA}}$, $\Delta\Phi_{\mathrm{CA}}$, and $\Delta\Phi_{\mathrm{DA}}$), which is understandable since the latter are derived from the fitting results and theoretically the mass slope $\gamma$ determines the lens mass distribution and thus determines the Fermat potential distribution (see Eq. (38) in \cite{Alessandro18}). Additional observational information, such as stellar velocity dispersion of the lens galaxy, can be possibly collected for providing complementary constraints on $\gamma$ and thus are helpful for reducing uncertainties of Fermat potential differences. More importantly, as suggested in Fig.~\ref{fig4}, the uncertainty of Fermat potential difference between two point sources is about 1\%. In time-delay cosmography, the concerned parameters, such as $H_0$ and $\Omega_k$, are inferred based on the combination of time delay and Fermat potential difference between each pair of images. For a quadruply lensed system, we find that the uncertainty from lens modeling on cosmological parameters ($H_0$ and $\Omega_k$) inference is 0.8\%. Here we choose a power-law model to fit the lens mass profile. In the literature, it has been noted that adopting the power-law mass distribution as a specific prior might lead to a potential bias due to mass-sheet degeneracy\cite{Schneider13, Xu15, Tagore17}. However, it also has been argued that such a bias could be reduced by carefully taking into account kinematic constraints and additional sources of systematic uncertainty \cite{Tagore17}. We want to point out here that this 0.8\% uncertainty level is valid when fitting the lens with a correct parameterized model (i.e., we generate the mock image with power-law lens mass profile and fit the image also with a power-law model).  Incorrectness of the lens model would lead to potential bias in the inference of $H_0$, where greater deviation of the models leading to more significant bias. For instance, when we use the ``Jaffe" model \cite{Jaffe83, Keeton01} to fit the mock lensing system shown in the left panel of Fig. \ref{fig3} where a power-law profile is considered, we even find $6\%-10\%$ bias in the inference of $H_0$. In practice, fortunately, high quality optical/IR image of the source-lens system could help us to avoid choosing obviously wrong models. To briefly demonstrate this, we use the power-law profile to simulate the lens arc but use the ``Jaffe'' mass model (i.e., a wrong model) to fit the mock arc. We find that when the exposure time is longer than 5000 seconds, the $\chi^2$ (Chisq) map for the ``Jaffe'' model starts to be prominent (see Fig.~\ref{fig5}) and the reduced $\chi^2$ values are much larger than the power-law ones (see Fig.~\ref{fig6}). This simple test demonstrates that the performance of different lens models could be distinguished when the quality of observed images is high enough (e.g., the exposure time is longer than 5000 seconds with the HST). For current available systems studied by the H0LiCOW program, the typical exposure time with the HST is $\sim 10^4$ seconds \cite{suyu17} . It is reasonable to expect that for each interesting lensed FRB system, extremely high quality images can be obtained from the HST or the near future JWST to distinguish among different lens models. Moreover, to mitigate such a potential bias, different parameterized models are often adopted so that a joint-consistent inference could be achieved. For example, in H0LiCOW IV \cite{wong16}, besides the power-law model, some other popular mass models were also adopted and resulted in consistent inferences with the power-law ones (see Fig. 9 therein). Overall, we conclude that for the near future lensed FRB systems of great interest, lens mass modeling would contribute an uncertainty at a 0.8\% level.
     
     The last ingredient of uncertainty contribution for time delay cosmography is the one from LOS environment modeling. The distribution of mass external to the lens, such as that associated with galaxies which are close in projection to the lens system along the LOS, affects the time delays between lensed images. An external convergence ($\kappa_{\rm {ext}}$) can be absorbed by the lens and source model leaving the fits to observables of a lens system (i.e., image positions, flux ratios for point sources, and the image shapes for extended sources) unchanged, but the predicted time delays altered by a factor of $(1-\kappa_{\rm {ext}})$. Consequently, the true $D_{\Delta t}$ is related to the modeled one via $D_{\Delta t}=D_{\Delta t}^{\rm {model}}/(1-\kappa_{\rm {ext}})$.  For the lens HE~0435-1223 \cite{wisotzki02}, by using various combinations of relevant informative weighing schemes for the galaxy counts \cite{fassnacht11} and ray-tracing through the Millennium Simulation \cite{springel05}, it was found that the most robust estimate of $\kappa_{\rm {ext}}$ has a median value $\kappa^{\rm {med}}_{\rm {ext}} = 0.004$ and a standard deviation $\sigma_\kappa = 0.025$, which corresponds to a 2.5\% uncertainty on the time delay distance \cite{cristian17}. More recently, using deep $r$-band images from Subaru-Suprime-Cam and an inpainting technique and Multi-Scale Entropy filtering algorithm, a weak gravitational lensing measurement of the external convergence along the line of sight to HE~0435-1223 has achieved $\kappa_{\rm {ext}} =- 0.012^{+0.020}_{-0.013}$, which corresponds to $\sim1.6\%$ uncertainty on the time delay distance \cite{Tihhonova17}. Furthermore, the distribution of $\kappa_{\rm {ext}}$ is robust to choices of weights, apertures, and flux limits, up to an impact of 0.5\% on the inferred time delay distance \cite{bonvin17}. Here we assume that for lensed FRBs systems, LOS environment contributes a systematic uncertainty at an averaged 2\% level to the inferred time delay distance, with expecting that every lensed FRB system is given enough attention by the community so that we can combine auxiliary follow-up data from facilities at different wavelengths and other available simulations with convergence maps. It is still worth noticing that larger LOS systematic uncertainty could lead to larger error bars of cosmological parameters. In addition to the distribution of mass external to the lens, the mass distribution in the outskirt of the lens halo, in which there are little optical light traces, might lead to errors in the inferred time delay distance at the percent level and it was shown that weak gravitational lensing and simulations may help to reduce these uncertainties \cite{Munoz17}.

\noindent
{\textbf{Distance sum rule}}
    In a homogeneous and isotropic universe with maximum symmetry, the spacetime is described by the FLRW metric
    \begin{equation}
     \textrm{d}s^2=-c^2\textrm{d}t^2+a^2(t)\left(\frac{\textrm{d}r^2}{1-Kr^2}+r^2\textrm{d}\Omega^2\right),
     \end{equation}
     where $a(t)$ is the scale factor and $K$ is a constant relating to the geometry of three dimensional space. In this metric,  the dimensionless comoving angular diameter distance of a source at redshift $z_s$ as observed at redshift $z_l$ is written as
    \begin{equation}
     d(z_l, z_s)=\frac{1}{\sqrt{\mid\Omega_k\mid}}S_K\bigg(\sqrt{\mid\Omega_k\mid}\int_{z_l}^{z_s}\frac{\textrm{d}x}{E(x)}\bigg),
     \end{equation}
     where
     \begin{equation}\label{dzlzs}
     S_K(X)=\left\{
     \begin{array}{ll}
     \displaystyle \sin(X) &~~~~~\Omega_k<0 \\
     \displaystyle X &~~~~~\Omega_k=0 \\
     \displaystyle \sinh(X) &~~~~~\Omega_k>0,
     \end{array}\right.
     \end{equation}
    $\Omega_k\equiv-K/H_0^2a_0^2$ ($a_0=a(0)$ is the present values of the scale factor and the Hubble parameter $H=\dot{a}/a$), and $E(z)\equiv H(z)/H_0$. In addition, we respectively denote $d(z)\equiv d(0,z)$, $d_l\equiv d(0,z_l)$, $d_s\equiv d(0,z_s)$, and $d_{ls}\equiv d(z_l,z_s)$. If the relation between the cosmic time $t$ and redshift $z$ is a single-valued function and $d'(z)>0$, these distances in the FLRW frame are connected via a simple sum rule~\cite{peebles93}
    \begin{equation}\label{sum1}
    d_{ls}=d_s\sqrt{1+\Omega_kd_l^2}-d_l\sqrt{1+\Omega_k d_s^2}.
    \end{equation}
     Apparently, the distances can be simply added together in a spatially flat universe. Notice that one has $d_s>d_l+d_{ls}$ or $d_s<d_l+d_{ls}$ for $\Omega_k>0$ or $\Omega_k<0$, respectively. Furthermore, Equation~(\ref{sum1}) can be rewritten as
     \begin{equation}\label{sum2}
     \frac{d_{ls}}{d_s}=\sqrt{1+\Omega_kd_l^2}-\frac{d_l}{d_s}\sqrt{1+\Omega_kd_s^2}.
     \end{equation}
     Recently, on the basis of Equation~(\ref{sum2}), a model-independent consistency test for the FLRW metric was discussed by comparing the distance ratios $d_{ls}/d_s$ measured from strongly lensed quasar systems with distances measured from SNe Ia observations\cite{rasanen15}. More recently, we upgraded the distance sum rule and rewrote it as \cite{liao17a}
     \begin{equation}\label{sum3}
     \frac{d_{ls}}{d_ld_s}=T(z_l)-T(z_s),      
     \end{equation}
     where
     \begin{equation}
     T(z)=\frac{1}{d(z)}\sqrt{1+\Omega_kd^2(z)}, 
     \end{equation}
     to test the FLRW metric and estimate the cosmic curvature with time delay distance measurements.

\noindent
{\textbf{Statistical analysis.}}
In order to estimate constraining power from 10 lensed FRB systems, we propagate the relative uncertainties of time delay ($\delta \Delta \tau=0$), Fermat potential difference ($\delta\Delta\Phi=0.8\%$), and line of sight contaminations ($\delta \kappa_{\mathrm{ext}}=2\%$) to the relative uncertainty of $D_{\mathrm{\Delta t}}$, and then to the (relative) uncertainties of cosmological parameters: $(\delta \Delta t, \delta\Delta\Phi, \delta \kappa_{\mathrm{ext}})\sim \delta D_{\mathrm{\Delta t}}\sim (\delta H_{\mathrm{0}},\sigma_{\Omega_k})$.
Then we perform Markov Chain Monte Carlo (MCMC) minimization of the following $\chi^2$ objective function:
\begin{equation}
\chi^2=\sum\limits_{i=1}^{10}{(D_{\mathrm{\Delta t},i}^{\mathrm{th}}(z_{\mathrm{d},i},z_{\mathrm{s},i};\mathbf{p})-D_{\mathrm{\Delta t},i}^{\mathrm{sim}})^2/\sigma^2_{ D_{\mathrm{\Delta t},i}}},
\end{equation}
where $D_{\mathrm{\Delta t}}^{\mathrm{th}}$ is the theoretical time delay distance in the assumed cosmological model or from the combination of distance sum rule and SNe Ia observations, while $D_{\mathrm{\Delta t}}^{\mathrm{sim}}$ is the corresponding simulated distance with its uncertainty is $\sigma_{D_{\mathrm{\Delta t},i}} = \delta D_{\mathrm{\Delta t},i} D_{\mathrm{\Delta t},i}$. $\mathbf{p}$ represents cosmological parameters ($H_{\mathrm{0}},\Omega_k$) to be constrained and they are sampled in ranges $H_{\mathrm{0}} \in [0, 150],~\Omega_k \in [-1, 1]$.

we perform 10000 simulations each containing 10 lensed FRB systems. For each dataset, we carry out the above-mentioned MCMC minimization to obtain the best-fit value of correponding parameters. Then we plot the probability distributions of the best-fit $H_0$ and $\Omega_k$ in Figure \ref{fig1} and Figure \ref{fig2}, respectively.

\noindent
{\textbf{Data availability.}} The data that support the findings of this study are available from the corresponding author upon request.

\begin{addendum}
\item [Acknowledgements] We would like to thank Liang Dai, Sherry H. Suyu, and Zong-Hong Zhu for reading the manuscript and their constructive suggestions. We are indebted to Kai Liao for his great help in lens modeling. This work was supported by the Ministry of Science and Technology National Basic Science Program (Project 973) under Grants No. 2014CB845806, the National Natural Science Foundation of China under Grants Nos. 11505008, 11722324, 11603003, 11373014, and 11633001, the Strategic Priority Research Program of the Chinese Academy of Sciences, Grant No. XDB23040100. X. Ding acknowledges support by China Postdoc Grant No. 2017M622501.

\item[Author contributions]
Z.-X.L. and H.G. initiated the idea using accurate measurements of time delays between images of lensed FRBs as a powerful probe for precision cosmology and drafted the paper. 
B.Z. contributed in discussion about this idea and in writing the manuscript. 
X.-H. D. and Z.-X.L. contributed in discussions and simulations for lens modeling. 
G.-J.W. and Z.-X.L. contributed in calculations for cosmological parameters inferences.

\item[Competing interests]The authors declare no competing interests.

\item[Materials \& Correspondence] Correspondence and requests for materials should be addressed to H.G. (gaohe@bnu.edu.cn).

\end{addendum}

\end{document}